\documentclass[10pt,onecolumn,twoside]{IEEEtran} 
\usepackage{graphics}
\usepackage{graphicx}
\usepackage{epsfig,amssymb}
\usepackage{amsmath}
\usepackage{mathrsfs}
\usepackage{color}
\usepackage{mathtools}
\usepackage{array}
\usepackage{amsfonts,booktabs}
\usepackage{lipsum,amsmath,multicol}
\usepackage{graphicx}
\usepackage{subfigure}
\usepackage{times}
\usepackage{pdfsync}
\usepackage{pgfplots}
\usepackage{pgfplotstable}
\usepackage[subnum]{cases}
\usepackage{filecontents}
\usepackage{multirow}
\usepackage{algorithm}
\usepackage{algpseudocode}
\usepackage{float}
\usepackage{setspace}
\newtheorem{theorem}{Theorem}
\newtheorem{lemma}{Lemma}

\newcommand{\beq}{\begin{equation}}
\newcommand{\eeq}{\end{equation}}
\newcommand{\beqa}{\begin{eqnarray}}
\newcommand{\eeqa}{\end{eqnarray}}

\newcommand{\paren}[1]{\left(#1\right)}
\newcommand{\sqparen}[1]{\left[#1\right]}

\newcommand{\field}[1]{\ensuremath{\mathbb{#1}}}
\newcommand{\abs}[1]{\left|#1\right|} 
\newcommand{\N}{\ensuremath{\field{N}}} 
\newcommand{\R}{\ensuremath{\field{R}}} 
\newcommand{\PRP}[1]{\ensuremath{\mathsf{Pr}\left(#1\right)}} 
\newcommand{\ES}[1]{\ensuremath{\mathsf{E}\left[#1 \right]}} 
\newcommand{\ENT}[1]{\ensuremath{\mathsf{h}\left[#1 \right]}} 
\newcommand{\e}[1]{\ensuremath{{\rm e}^{#1}}} 

\newcommand{\DE}[1]{\ensuremath{\mathsf{h}\sqparen{#1}}}
\newcommand{\DSE}[1]{\ensuremath{\mathsf{H}\sqparen{#1}}}
\newcommand{\DEP}[1]{\ensuremath{\mathsf{N}_{\rm d}\left[#1 \right]}} 
\newcommand{\CEP}[1]{\ensuremath{\mathsf{N}_{\rm c}\left[#1 \right]}} 
\begin{document}

\title{An Entropy Power Inequality for Discrete Random Variables}
\author{ Ehsan Nekouei, Mikael Skoglund and Karl H. Johansson
\thanks{School of electrical engineering and computer science, KTH Royal Institute of Technology,   Stockholm, Sweden. \{nekouei,skoglund,kallej\}@kth.se. This work is supported by the Knut and Alice Wallenberg Foundation, the
Swedish Foundation for Strategic Research and the Swedish Research Council.}}
\maketitle
\thispagestyle{empty}

\begin{abstract}
Let $\DEP{X}=\frac{1}{2\pi \e{}}\e{2\DSE{X}}$ denote the entropy power of the discrete random variable $X$ where $\DSE{X}$ denotes the discrete entropy of $X$. In this paper, we show that for two independent discrete random variables $X$ and $Y$, the entropy power inequality $\DEP{X}+\DEP{Y}\leq 2 \DEP{X+Y}$ holds and it can be  tight. The basic idea behind the proof is to perturb the discrete random variables using suitably designed continuous random variables. Then, the continuous entropy power inequality is applied to the sum of the perturbed random variables and the resulting lower bound is optimized.    
\end{abstract}

\section{Introduction}

The \emph{continuous} entropy power inequality \cite{Shannon48}, \cite{STAM59}, \cite{Blachman65} asserts that for two independent absolutely continuous random variables (rvs) $U$ and $V$, the following inequality holds
\begin{align}\label{Eq: CEP}
\CEP{U}+\CEP{V}\leq \CEP{U+V},
\end{align}
where $\CEP{\cdot}=\frac{1}{2\pi e{}}\e{2\ENT{\cdot}}$ and $\ENT{\cdot}$ denote the continuous entropy power and the differential entropy functionals, respectively. In the information theory literature, substantial efforts have been dedicated to obtaining an analogue of \eqref{Eq: CEP} for discrete rvs. In general, the discrete counterpart of  \eqref{Eq: CEP}, where the differential entropy is replaced by the discrete entropy, does not hold for discrete rvs.  Classes of discrete rvs which satisfy the discrete version of \eqref{Eq: CEP} have been studied in the literature. Let $B\paren{n,p}$ denote a binomial distribution with $n$ trials and success probability $p$. Harremo\"es and Vignat \cite{HV2003} showed that the discrete version of \eqref{Eq: CEP} holds for two binomial rvs distributed according to $B\paren{n,p}$ and $B\paren{m,p}$ with $p=\frac{1}{2}$ and $m,n\in\N$. Sharma \emph{et al.,} \cite{SDM11} proved that this result holds for $p\in\paren{0,1}$ when $m$ and $n$ are sufficiently large.

 The authors of \cite{WM15} showed that the discrete version of \eqref{Eq: CEP} holds for two independent and uniformly distributed rvs. A variant of the entropy power inequality for ultra log-concave discrete rvs has been derived in \cite{JY10} using R\'enyi's thinning operation. It worth mentioning that lower bounds on the entropy of a sum of independent discrete rvs have been investigated extensively in the literature. The interested reader is referred to \cite{HAT14}, \cite{JA14} and references therein for more information on this line of research.  

In this paper, we derive a discrete entropy power inequality, which is analogous to the continuous entropy power inequality and holds for the sum of two arbitrarily distributed, independent discrete rvs. More specifically, it is shown that for two independent discrete rvs $X$ and $Y$, we have 
\begin{align}
\DEP{X}+\DEP{Y}\leq 2 \DEP{X+Y}\nonumber
\end{align}
regardless of their distributions, where $\DEP{\cdot}=\frac{1}{2\pi \e{}}\e{2\DSE{\cdot}}$ and $\DSE{\cdot}$  denote the {discrete} entropy power and the {discrete} entropy, respectively.
\subsection{Notation and Organization of The Paper}
Let $V$ denote a generic continuous random variable taking values on $\R$. The differential entropy of $V$ and its (continuous) entropy power are defined as 
\begin{align}
\DE{V}&\coloneqq-\int p_V\paren{x}\log p_V\paren{x} dx,\nonumber\\
\CEP{V}&\coloneqq\frac{1}{2\pi e{}}\e{2\ENT{V}},\nonumber
\end{align}
 where $p_V\paren{x}$ denotes the probability density function (pdf) of $V$. For a generic discrete random variable $X$, its discrete entropy and entropy power are defined as 
 \begin{align}
 \DSE{X}&\coloneqq-\sum_i\PRP{X=x_i}\log\PRP{X=x_i},\nonumber\\
 \DEP{X}&\coloneqq\frac{1}{2\pi e{}}\e{2\DSE{X}}.\nonumber
 \end{align}
  
  The rest of this paper is organized as follows. Next section presents our main result along with the key steps of its proof. Detailed proofs of the steps are presented in Section \ref{Sec: Proof}. 
\section{The Main Result}
The following theorem establishes an entropy power inequality for the sum of two independent discrete rvs.
\begin{theorem}\label{Theo: EPI}
Consider two independent discrete rvs $X$ and $Y$. Then, we have 
\begin{align}\label{Eq: DEPI}
\DEP{X}+\DEP{Y}\leq 2 \DEP{X+Y}.
\end{align} 
Moreover, the equality is achieved when the ``effective" support sets of $X$ and $Y$ are singletons.  
\end{theorem}

Theorem \ref{Theo: EPI} establishes an upper bound on the sum of entropy powers of two independent discrete rvs. According to this result, the sum of the entropy powers of two independent discrete rvs is always less than twice of the entropy power of their sum. Also, the inequality is tight when  each rv only takes one value from its support set with probability one. Note that the difference between the two sides of \eqref{Eq: DEPI} becomes small when the probability mass function of each rv is highly concentrated around one element of its support set. 

\subsection{Proof of Theorem \ref{Theo: EPI}}
   The proof of Theorem \ref{Theo: EPI} relies on $1)$ perturbing the discrete rvs by carefully chosen continuous rvs, $2)$ applying the continuous entropy power inequality to the sum of perturbed rvs, and $3)$  optimizing the lower bound obtained in step $2$. In this subsection, Theorem \ref{Theo: EPI} is proved using four key lemmas.  
   
   Let $M$ denote a discrete rv taking values in $\left\{m_1,\dots,m_k\right\}$ and $\alpha_m$ denote the minimum spacing between its atoms, \emph{i.e,} $\alpha_m=\min_{i\neq j}\abs{m_i-m_j}$. Also, let $T$ denote a real-valued rv, independent of $M$, with $\abs{T}<\frac{\alpha_m}{2}$ almost surely (a.s.). We assume that $T$ is absolutely continuous with respect to the Lebesgue measure on the real line  and has finite differential entropy. 

 The following lemma derives an expression for the differential entropy of $M+T$. Its proof is presented in Subsection \ref{Subsec: Lem_DFE}.
\begin{lemma}\label{Lem: DFE}
 The differential entropy of $M+T$ can be written as 
\begin{align}\label{Eq: Ent-Eq}
\DE{M+T}=\DSE{M}+\DE{T},
\end{align}
where $\DE{\cdot}$ and $\DSE{\cdot}$ denote the differential entropy and the discrete entropy, respectively. 
\end{lemma}

 Let $X$ and $Y$ denote independent discrete rvs, and $Z$ denote their sum. Let $\alpha_x$, $\alpha_y$ and $\alpha_z$ denote the minimum spacing of $X$, $Y$ and $Z$, respectively. Next lemma derives an upper bound on the minimum spacing of $Z$. The proof of this result is straightforward and is skipped.
\begin{lemma}\label{Lem: Spac}
We have $\alpha_z\leq \min\paren{\alpha_x,\alpha_y}$.
\end{lemma}
According to this lemma, the minimum spacing between the atoms of $Z$ is not larger than those of $X$ and $Y$. 

 Let $W_1$ and $W_2$ be independent and identically distributed (iid) absolutely continuous rvs which are independent of $X$ and $Y$; and take values in $\paren{-\frac{\alpha_z}{4},\frac{\alpha_z}{4}}$. Let $p\paren{x}$ denote the common probability density function (pdf) of $W_1$ and $W_2$ (with respect to the Lebesgue measure on the real line) and assume it has finite differential entropy.  Consider the rvs $X+W_1$ and $Y+W_2$ which are obtained by perturbing $X$ and $Y$ using $W_1$ and $W_2$.  From Lemmas \ref{Lem: DFE} and \ref{Lem: Spac}, we have
\begin{align}\label{Eq: Ent-Eq-XY}
\DE{X+W_1}&=\DSE{X}+\DE{W_1}\nonumber\\
\DE{Y+W_2}&=\DSE{Y}+\DE{W_2}.
\end{align} 
 Moreover, using Lemma \ref{Lem: DFE} and the fact that $\abs{W_1+W_2}<\frac{\alpha_z}{2}$ a.s., we have 
 \begin{align}\label{Eq: Ent-Eq-Z}
 \DE{X+W_1+Y+W_2}=\DSE{X+Y}+\DE{W_1+W_2}.
 \end{align}  

The equalities \eqref{Eq: Ent-Eq-XY} and \eqref{Eq: Ent-Eq-Z} are used to establish an inequality on the entropy power of $X+Y$ in Lemma \ref{Lem: Sup_LB}. This lemma is proved in Subsection \ref{Subsec: Lem_Sup_LB} by applying the continuous entropy power inequality to the sum of the perturbed rvs $X+W_1$ and $Y+W_2$. 
\begin{lemma}\label{Lem: Sup_LB}
Let $\Lambda$ denote the set of pdfs defined on $\paren{-\frac{\alpha_z}{4},\frac{\alpha_z}{4}}$ and have finite differential entropies. Then, we have 
\begin{align}
\frac{\DEP{X+Y}}{\DEP{X}+\DEP{Y}}\geq \sup_{p\paren{x}\in \Lambda}\frac{\e{2\DE{W_1}}}{\e{2\DE{W_1+W_2}}},\nonumber
\end{align}
where $W_1$ and $W_2$ are two independent absolutely continuous rvs with pdf $p\paren{x}\in\Lambda$. 
\end{lemma}

Next lemma characterizes the lower bound in Lemma \ref{Lem: Sup_LB}. The proof of this lemma is relegated to  Subsection \ref{Subsec: Sup_Eq}. 
\begin{lemma}\label{Lem: Sup_eq}

\begin{align}
\sup_{p\paren{x}\in \Lambda}\frac{\e{2\DE{W_1}}}{\e{2\DE{W_1+W_2}}}=\frac{1}{2}.\nonumber
\end{align}
\end{lemma}

 The proof of Theorem \ref{Theo: EPI} follows from Lemmas \ref{Lem: Sup_LB} and \ref{Lem: Sup_eq}.
 \section{Proofs of Lemmas}\label{Sec: Proof}
  \subsection{Proof of Lemma \ref{Lem: DFE}}\label{Subsec: Lem_DFE}
 
 Let $P_T\paren{x}$ denote the pdf of $T$. Then, the pdf of $M+T$ can be written as $\sum_i\PRP{M=m_i}P_{T}\paren{x-m_i}$. The assumption $\abs{T}<\frac{\alpha_m}{2}$ implies that the size of the support set of $T$ is less than the minimum spacing of $M$. This observation implies  that the pdf of $M+T$ is composed of $k$  non-overlapping components. Using the definition of the differential entropy, we have
 \begin{align}
 \DE{M+T} 
 &=-\int \sum_{i}\PRP{M=m_i}P_{T}\paren{x-m_i}\log \sum_i\PRP{M=m_i}P_{T}\paren{x-m_i}dx\nonumber\\
 &\stackrel{(a)}{=}-\sum_i\int \PRP{M=m_i}P_T\paren{x-m_i}\log \PRP{M=m_i}P_T\paren{x-m_i}dx\nonumber\\
 &=-\sum_i\PRP{M=m_i}\log \PRP{M=m_i}\int P_T\paren{x-m_i}dx-\sum_i\PRP{M=m_i}\int P_T\paren{x-m_i}\log P_T\paren{x-m_i}dx\nonumber\\
 &\stackrel{(b)}{=}-\sum_i\PRP{M=m_i}\log \PRP{M=m_i}-\int P_T\paren{x}\log P_T\paren{x}dx\nonumber\\
 &=\DSE{M}+\DE{T},\nonumber
 \end{align}
 where $(a)$ follows from the fact that the components of the pdf of $M+T$ are non-overlapping and $(b)$ from the fact that the differential entropy is shift-invariant. 
 \subsection{Proof of Lemma \ref{Lem: Sup_LB}}\label{Subsec: Lem_Sup_LB}
 Using the entropy power inequality for continuous rvs \cite{Shannon48}, we have 
 \begin{align}
 1&\geq \frac{\CEP{X+W_1}+\CEP{Y+W_2}}{\CEP{X+W_1+Y+W_2}}\nonumber\\
 &=\frac{\frac{1}{2\pi\e{}}\e{2\DE{X+W_1}}+\frac{1}{2\pi\e{}}\e{2\DE{Y+W_2}}}{\frac{1}{2\pi\e{}}\e{2\DE{X+W_1+Y+W_2}}}\nonumber\\
 &\stackrel{(a)}{=}\frac{\frac{1}{2\pi\e{}}\e{2\DSE{X}}\e{2\DE{W_1}}+\frac{1}{2\pi\e{}}\e{2\DSE{Y}}\e{2\DE{W_2}}}{\frac{1}{2\pi\e{}}\e{2\DSE{X+Y}}\e{2\DE{W_1+W_2}}}\nonumber\\
 &\stackrel{(b)}{=}\frac{\e{2\DE{W_1}}}{\e{2\DE{W_1+W_2}}}\frac{\DEP{X}+\DEP{Y}}{\DEP{X+Y}},\nonumber
 \end{align}
 where $(a)$ follows from equalities \eqref{Eq: Ent-Eq-XY} and \eqref{Eq: Ent-Eq-Z} and $(b)$ follows from the fact that $W_1$ and $W_2$ are identically distributed. Hence, we have 
 \begin{align}\label{Eq: Ineq-1}
 \frac{\DEP{X+Y}}{\DEP{X}+\DEP{Y}}\geq \frac{\e{2\DE{W_1}}}{\e{2\DE{W_1+W_2}}}.
 \end{align}
 
Inequality \eqref{Eq: Ineq-1} holds for any pdf defined on $\paren{-\frac{\alpha_z}{2},\frac{\alpha_z}{2}}$ with a finite differential entropy.  Thus, we have 
 \begin{align}
 \frac{\DEP{X+Y}}{\DEP{X}+\DEP{Y}}\geq \sup_{p\paren{x}\in \Lambda}\frac{\e{2\DE{W_1}}}{\e{2\DE{W_1+W_2}}}.\nonumber
 \end{align}
 \subsection{Proof of Lemma \ref{Lem: Sup_eq}}\label{Subsec: Sup_Eq}
  Using the entropy power inequality for continuous rvs, we have 
 \begin{align}
 \frac{\e{2\DE{W_1}}}{\e{2\DE{W_1+W_2}}}\leq \frac{1}{2}\nonumber
 \end{align}
 for all independent and identically distributed rvs $W_1$ and $W_2$ with the common pdf in $\Lambda$. Thus, we have 
 \begin{align}
 \sup_{p\paren{x}\in \Lambda}\frac{\e{2\DE{W_1}}}{\e{2\DE{W_1+W_2}}}\leq \frac{1}{2}.\nonumber
 \end{align}
   To show the other direction, let $N\paren{0,\sigma^2}$ denote the pdf of a Gaussian rv with zero mean and variance $\sigma^2$. Let $p_\sigma\paren{x}$ denote the pdf obtained by truncating $N\paren{0,\sigma^2}$ outside $\paren{-\frac{\alpha_z}{4},\frac{\alpha_z}{4}}$, \emph{i.e.,}  
   \begin{align}
   p_\sigma\paren{x}=\left\{
   \begin{array}{cc}
  \frac{K\paren{\sigma}}{\sqrt{2\pi}\sigma}\e{-\frac{x^2}{2\sigma^2}}&x\in\paren{-\frac{\alpha_z}{4},\frac{\alpha_z}{4}}\nonumber\\
  0& \text{o.w.},\nonumber
   \end{array}
   \right.
   \end{align}
   where $K\paren{\sigma}=\paren{\int_{-\frac{\alpha_z}{4}}^{\frac{\alpha_z}{4}} \frac{1}{\sqrt{2\pi}\sigma}\e{-\frac{x^2}{2\sigma^2}}dx}^{-1}$ is the normalizing factor. Let $W_1^\sigma$ and $W_2^\sigma$ be two independent rvs distributed according to $p_\sigma\paren{x}$. Then, we have 
    \begin{align}\label{Eq: Sup-LB}
    \sup_{p\paren{x}\in \Lambda}\frac{\e{2\DE{W_1}}}{\e{2\DE{W_1+W_2}}}&\stackrel{(a)}{\geq} \frac{\e{2\DE{W_1^\sigma}}}{\e{2\DE{W_1^\sigma+W_2^\sigma}}}\nonumber\\
    &\stackrel{(b)}{\geq}\frac{\e{2\DE{W_1^\sigma}}}{\e{2\times \frac{1}{2}\log\paren{2\pi\e{}\ES{\paren{W_1^\sigma+W_2^\sigma}^2}}}}\nonumber\\
    &\geq\frac{\e{2\DE{W_1^\sigma}}}{2\pi\e{}\ES{\paren{W_1^\sigma+W_2^\sigma}^2}},
    \end{align}
    where $(a)$ follows from the fact that $  p_\sigma\paren{x}$ belongs to $\Lambda$ and $(b)$ follows from the entropy maximizing property of Gaussian distributions. The variance of $W_1^\sigma+W_2^\sigma$ can be upper bounded as 
    \begin{align}\label{Eq: M-Exp}
    \ES{\paren{W_1^\sigma+W_2^\sigma}^2}&=2\ES{\paren{W_1^\sigma}^2}\nonumber\\
    &=2\int_{-\frac{\alpha_z}{4}}^{\frac{\alpha_z}{4}}x^2 \frac{K\paren{\sigma}}{\sqrt{2\pi}\sigma}\e{-\frac{x^2}{2\sigma^2}}dx\nonumber\\
    &\leq 2K\paren{\sigma}\int_{-\infty}^{\infty}\frac{x^2 }{\sqrt{2\pi}\sigma}\e{-\frac{x^2}{2\sigma^2}}dx\nonumber\\
      &=2K\paren{\sigma}\sigma^2.    
    \end{align}
    Moreover, the differential entropy of ${W_1^\sigma}$ can be written as 
    \begin{align}\label{Eq: DE-Exp}
    \DE{W_1^\sigma}&=-\int_{-\frac{\alpha_z}{4}}^{\frac{\alpha_z}{4}}p_\sigma\paren{x}\log\frac{K\paren{\sigma}}{\sqrt{2\pi}\sigma}\e{-\frac{1}{2\sigma^2}x^2}dx\nonumber\\
   & =-\log K\paren{\sigma}-K\paren{\sigma}\int_{-\frac{\alpha_z}{4}}^{\frac{\alpha_z}{4}}\frac{1}{\sqrt{2\pi}\sigma}\e{-\frac{1}{2\sigma^2}x^2}\log\frac{1}{\sqrt{2\pi}\sigma}\e{-\frac{1}{2\sigma^2}x^2}dx\nonumber\\
    & =-\log K\paren{\sigma}-K\paren{\sigma}\left[\int_{-\infty}^{\infty}\frac{1}{\sqrt{2\pi}\sigma}\e{-\frac{1}{2\sigma^2}x^2}\log\frac{1}{\sqrt{2\pi}\sigma}\e{-\frac{1}{2\sigma^2}x^2}dx-2\int_{\frac{\alpha_z}{4}}^{\infty}\frac{1}{\sqrt{2\pi}\sigma}\e{-\frac{1}{2\sigma^2}x^2}\log\frac{1}{\sqrt{2\pi}\sigma}\e{-\frac{1}{2\sigma^2}x^2}dx\right]\nonumber\\
       & =-\log K\paren{\sigma}-K\paren{\sigma}\left[-\frac{1}{2}\log\paren{2\pi\e{}\sigma^2}-2\int_{\frac{\alpha_z}{4\sigma}}^{\infty}\frac{1}{\sqrt{2\pi}}\e{-\frac{1}{2}x^2}\log\frac{1}{\sqrt{2\pi}\sigma}\e{-\frac{1}{2}x^2}dx\right]\nonumber\\
        & =-\log K\paren{\sigma}-K\paren{\sigma}\left[-\frac{1}{2}\log\paren{2\pi\e{}\sigma^2}+2\underset{\eta\paren{\sigma}}{\underbrace{\log\sqrt{2\pi}\sigma\int_{\frac{\alpha_z}{4\sigma}}^{\infty}\frac{1}{\sqrt{2\pi}}\e{-\frac{1}{2}x^2}dx}}+\underset{\Phi\paren{\sigma}}{\underbrace{\int_{\frac{\alpha_z}{4\sigma}}^{\infty}\frac{x^2}{\sqrt{2\pi}}\e{-\frac{1}{2}x^2}dx}}\right]\nonumber\\
        & =-\log K\paren{\sigma}-K\paren{\sigma}\left[-\frac{1}{2}\log\paren{2\pi\e{}\sigma^2}+2\eta\paren{\sigma}+\Phi\paren{\sigma}\right].
    \end{align}
    
  Using \eqref{Eq: M-Exp} and \eqref{Eq: DE-Exp}, we have 
    \begin{align}
    \frac{\e{2\DE{W_1^\sigma}}}{2\pi\e{}\ES{\paren{W_1^\sigma+W_2^\sigma}^2}}&\geq \frac{\e{\log\paren{2\pi\e{}\sigma^2}}\e{\log\paren{2\pi\e{}\sigma^2}\left[K\paren{\sigma}-1\right]}\e{-2\log K\paren{\sigma}-2K\paren{\sigma}\left[2\eta\paren{\sigma}+\Phi\paren{\sigma}\right]}}{4\pi\e{}K\paren{\sigma}\sigma^2}\nonumber\\
    &=\frac{2\pi\e{}\sigma^2\e{\log\paren{2\pi\e{}\sigma^2}\left[K\paren{\sigma}-1\right]}\e{-2\log K\paren{\sigma}-2K\paren{\sigma}\left[2\eta\paren{\sigma}+\Phi\paren{\sigma}\right]}}{4\pi\e{}K\paren{\sigma}\sigma^2}\nonumber\\
       &=\frac{1}{2}\frac{\e{\log\paren{2\pi\e{}\sigma^2}\left[K\paren{\sigma}-1\right]}\e{-2\log K\paren{\sigma}-2K\paren{\sigma}\left[2\eta\paren{\sigma}+\Phi\paren{\sigma}\right]}}{K\paren{\sigma}}\nonumber\\
       &\coloneqq\frac{1}{2}F\paren{\sigma}.\nonumber
    \end{align}
    
      Note that $\lim_{\sigma\downarrow 0}K\paren{\sigma}=1$ and $\lim_{\sigma\downarrow 0}\Phi\paren{\sigma}=0$.  The term $
       \abs{\eta\paren{\sigma}}$ can be upper bounded as 
       \begin{align}
       \abs{\eta\paren{\sigma}}&=\abs{\log\sqrt{2\pi}\sigma}\int_{\frac{\alpha_z}{4\sigma}}^{\infty}\frac{1}{\sqrt{2\pi}}\e{-\frac{1}{2}x^2}dx\nonumber\\
       &\stackrel{(a)}{\leq} \abs{\log\sqrt{2\pi}\sigma}\e{-\frac{1}{2}\paren{\frac{\alpha_z}{4\sigma}}^2},\nonumber
       \end{align} 
      where $(a)$ follows from the fact that $\int_{x}^{\infty}\frac{1}{\sqrt{2\pi}}\e{-\frac{1}{2}x^2}dx\leq \e{-\frac{x^2}{2}}$ for $x>0$ \cite{Verdu98}. Thus, we have  $\lim_{\sigma\downarrow 0}\eta\paren{\sigma}=0$. The term $K\paren{\sigma}-1$ can be written as 
       \begin{align}
       K\paren{\sigma}-1&=\frac{1-\int_{-\frac{\alpha_z}{4}}^{\frac{\alpha_z}{4}} \frac{1}{\sqrt{2\pi}\sigma}\e{-\frac{x^2}{2\sigma^2}}dx}{\int_{-\frac{\alpha_z}{4}}^{\frac{\alpha_z}{4}} \frac{1}{\sqrt{2\pi}\sigma}\e{-\frac{x^2}{2\sigma^2}}dx}\nonumber\\
       &=\frac{2\int_{\frac{\alpha_z}{4\sigma}}^{\infty} \frac{1}{\sqrt{2\pi}}\e{-\frac{x^2}{2}}dx}{\int_{-\frac{\alpha_z}{4}}^{\frac{\alpha_z}{4}} \frac{1}{\sqrt{2\pi}\sigma}\e{-\frac{x^2}{2\sigma^2}}dx}\nonumber\\
          &\leq \frac{2\e{-\frac{1}{2}\paren{\frac{\alpha_z}{2\sigma}}^2}}{\int_{-\frac{\alpha_z}{4}}^{\frac{\alpha_z}{4}} \frac{1}{\sqrt{2\pi}\sigma}\e{-\frac{x^2}{2\sigma^2}}dx},\nonumber
       \end{align}
       which implies that $\lim_{\sigma\downarrow 0}\log\paren{2\pi\e{}\sigma^2}\left[K\paren{\sigma}-1\right]=0$. Thus, we have $\lim_{\sigma\downarrow 0}F\paren{\sigma}=1$.
       
       For a given $\epsilon>0$, we can find $\sigma_0$ small enough such that $F\paren{\sigma_0}\geq 1-\epsilon$ and $p_{\sigma_0}\paren{x}\in\Lambda$. Thus, we have  
          \begin{align}
           \sup_{p\paren{x}\in \Lambda}\frac{\e{2\DE{W_1}}}{\e{2\DE{W_1+W_2}}}&\geq  \frac{\e{2\DE{W_1^{\sigma_0}}}}{2\pi\e{}\ES{\paren{W_1^{\sigma_0}+W_2^{\sigma_0}}^2}}\nonumber\\
         &\geq \frac{1}{2}-\frac{\epsilon}{2}\nonumber
          \end{align}
          for $\sigma_0$ sufficiently small. The desired result follows from the fact that $\epsilon>0$ is arbitrary.  
  \bibliographystyle{IEEEtran}
  \bibliography{Ref}
  
\end{document}